\documentclass[aps,prl,showpacs]{revtex4} 
\usepackage{subfig}
\usepackage{float}
\usepackage{epsfig}
\usepackage[usenames,dvipsnames]{color}
\newcommand{\be}{\begin{equation}}
\newcommand{\ee}{\end{equation}}
\newcommand{\ba}{\begin{eqnarray}}
\newcommand{\ea}{\end{eqnarray}}

\newcommand{\bwt}{ \begin{widetext}}
\newcommand{\ewt}{ \end{widetext}}

\newcommand{\beq}{\begin{equation}}
\newcommand{\eeq}{\end{equation}}

\newcommand{\half}{\frac 1 2 }

\newcommand{\call}{{\cal L}}
\begin{document}
\title{Effective metric in nonlinear scalar field theories}
\author{\ E. Goulart}
\email{egoulart2@gmail.com}
\affiliation{Centro Brasileiro de Pesquisas Fisicas, Rua
Xavier Sigaud, 150, CEP 22290-180, Rio de Janeiro, Brazil.
}
\author{\ Santiago Esteban Perez Bergliaffa}
\email{sepbergliaffa@gmail.com}
\affiliation{Departamento de F\'{\i}sica Te\'{o}rica,
Instituto de F\'{\i}sica, Universidade do Estado de Rio de
Janeiro, CEP 20550-013, Rio de Janeiro, Brazil.}
\vspace{.5cm}
\begin{abstract}
We discuss several features of the propagation of perturbations in nonlinear scalar field theories using the effective metric. It is shown that the effective metric can be classified according
to whether the gradient of the scalar field is timelike, null, or spacelike, and this classification is illustrated with two examples. We shall also show that different signatures for the effective metric are allowed. 
\end{abstract}
\pacs{pacs}
\maketitle

\section{Introduction}

It has been known for some time that the propagation of the 
excitations of nonlinear field theories in a given background is governed by an effective metric that depends 
on the background field configuration and on the details of the non-linear dynamics (see \cite{mattlr} for a review).  
The perturbations in systems such as moving fluids, Bose-Einstein
condensates, superfluids, and nonlinear electromagnetism, to mention a few
\footnote{See \cite{mattlr}
for a complete list.} evolve in a curved effective spacetime different from the background geometry (which can even 
be Newtonian). Among other developments, the effective metric led to the construction of
analog models of gravity, which imitate the kinematical properties of gravitational fields
\footnote{See however ref.\cite{marioerico} for a case in which it is possible to go beyond
the kinematics.}, and to insights on the possibility of understanding gravity as an emergent phenomenon
\cite{emergent}. 

Being dependent on the background field (as well as on the details of the nonlinear theory), it is logical to expect that the effective metric could be classified in different types, according to the  background. We have shown that this is indeed the case in nonlinear theories for the electromagnetic field in \cite{nosso}, where we showed that only two types of effective metrics are possible in this case. This was achieved by using the dependence of the effective metric on the
energy-momentum tensor of the background along with the Segr\`e classification of the latter.
Here we continue along this line with the analysis of the effective geometry in the case of nonlinear scalar field 
theories. These theories have been intensively studied in the last ten years to account for inflation \cite{piconmu} (and alternatives to it, see 
for instance \cite{bessada}, \cite{hu}), 
to model dark energy
\cite{kessence}, and to unify dark matter and dark energy \cite{unify}.
These applications paved the way for the discussion of other issues, such as 
topological defects \cite{babi}, collapse \cite{collapse}, the possibility of signals escaping from the interior of a black hole \cite{vik}, and the formation of haloes
of dark energy \cite{picon}. Typically, in all these works a noncanonical kinetic term is involved, which also plays an important role in
ghost condensation \cite{ghost}
and Galileon fields \cite{galileon}.

After the introduction, we show
that the different possibilities for the effective metric are specified by the character of the 
gradient of the field (i.e. spacelike, timelike, or null). Some examples of solutions of nonlinear scalar field theories already studied in the literature are also examined from the point 
of view of this classification.
In all these sections, it was assumed that there is 
a lower bound for the Hamiltonian of the system (thus guaranteeing the stability of the theory), which translates in $\call_{W}>0$ (see for instance
\cite{aha}). 
It was also assumed that the field equations are hiperbolic, which entails that the Cauchy problem is well-posed in the sense of Hadamard (that is,  when a unique solution exists and depends continuously upon the data). Since, as we shall see below, it is the effective metric that sets the type of partial differential equation describing the background field, in Sec.\ref{effe} we shall lift 
the requirement of hyperbolicity, and discuss what types of equation are possible, with the corresponding signature. We close with some comments.

\section{Scalar fields in curved spacetimes}
\label{intro}
Let us begin by recalling some properties of the propagation of
a test real scalar field in a given Riemannian and globally hyperbolic spacetime $\mathcal{M}$, endowed with a metric tensor $g_{\mu\nu}$, with signature $(+,-,-,-)$. 
From the action
\begin{equation}\label{action}
S=\int\sqrt{-g}\;\call (\phi,W)
\end{equation}
where the lagrangian density $\call(\phi,W)$ is a real function of the field $\phi $ and $W\equiv\nabla_{\mu}\phi\nabla^{\mu}\phi$,
we obtain the equation of motion $
\nabla_{\mu}\left(\call_{W}\nabla^{\mu}\phi\right)=\frac{1}{2}\call_{\phi}
$,
with $\call_{x}\equiv\partial\call/\partial x$. This equation can be written as 
\begin{equation}\label{eom}
\left\{\call_Wg^{\mu\nu}+2\call_{WW}\phi^{\mu\nu}\right\}\nabla_{\mu} \nabla_{\nu}\phi+\call_{\phi W}\phi^{\alpha}_{\phantom a\alpha}-\frac{1}{2}\call_{\phi}=0,\label{pde}
\end{equation}
where $\phi^{\mu\nu}\equiv\nabla^{\mu}\phi\nabla^{\nu}\phi$ to simplify notation. As the principal part of the equation is linear in the second derivatives it constitutes a second order quasi-linear partial differential equation. 
The propagation of small excitations of the field around a given known background solution can be studied by means of the 
eikonal approximation for the field, that is
\begin{equation}
\phi(x)=\phi_{0}(x)+\epsilon(x){\rm exp}[-i\Sigma(x)],
\end{equation}
where $\phi_{0}(x)$ is an exact solution of the equation of motion (\ref{eom}), $\epsilon(x)$ a slowly varying infinitesimal amplitude and $\Sigma(x)$ a rapidly varying phase. Defining $k_{\alpha}\equiv\nabla_{\alpha}\Sigma$ and retaining only the phase gradients, we obtain the well known result that the field excitations satisfies a dispersion relation of the form \cite{mm}
\begin{equation}
\hat g^{\mu\nu}k_{\mu}k_{\nu}=0,\label{disp}
\end{equation}
where $\hat g^{\mu\nu}$ 
is given by
\begin{equation}
\hat g^{\mu\nu}\equiv\left.(\call_{W}g^{\mu\nu}+2\call_{WW}\phi^{\mu\nu})\right|_0,
\label{effmet}
\end{equation}
and the subindex 0 means that all the quantities are evaulated at a given background solution $\phi_0$. 

At a given point of spacetime, the set of all possible vectors $\left\{k_{\alpha}\right\}$ satisfying Eqn.(\ref{disp}) determines a three-dimensional surface with the topology of a cone. It is common to identify $\hat g^{\mu\nu}(x)$ as the contravariant components of an effective Riemannian metric in which the gradient $k_{\alpha}$ is a null vector. It is possible to show that the ``scalar rays'' are described by the null geodesics of this metric, while the same rays are accelerated curves with respect to the background spacetime $\mathcal{M}$. Let us analyze some basic properties of Eqn.(\ref{disp}), by taking $\hat g^{\mu\nu}$ as a field defined in the background $\mathcal{M}$ with metric $g^{\mu\nu}$. 
If ${\rm det}(\hat g^{\mu\nu})\neq0$, it is possible to define the inverse object $\hat g_{\mu\nu}$, such that
$
\hat g^{\mu\alpha}\hat g_{\alpha\nu}=\delta^{\mu}_{\phantom a\nu}.
$
A straightforward calculation shows that the inverse matrix $\hat g_{\mu\nu}$ is given by 
\begin{equation}\label{in1}
\hat g_{\mu\nu}=ag_{\mu\nu}+b\phi_{\mu\nu},
\end{equation}\label{in2}
where
\begin{equation}
a=\frac{1}{\call_{W}},\quad\quad\quad\quad b=-\frac{2\call_{WW}}{\call_{W}(\call_{W}+2\call_{WW}W)}.
\eeq
It is convenient also to define the object $q^{\mu}\equiv \hat g^{\mu\nu}k_{\nu}$, dual to the gradient $k_{\nu}$. From Eqn.(\ref{disp}) it is seen that $q^{\mu}$ and $k_{\mu}$ are orthogonal vectors \textit{i.e.} $q^{\mu}$ is tangent to the surfaces $\Sigma=const$. It also follows that
\begin{equation}\label{cones}
\hat g_{\mu\nu}q^{\mu}q^{\nu}=0.\label{disp2}
\end{equation}
Although relations (\ref{disp}) and (\ref{disp2}) seem to encompass the same properties, the objects $k^{\mu}$ and $q^{\mu}$ have complementary physical/geometrical meaning. In fact, equations (\ref{disp}) and (\ref{disp2}), determine two types of conoid structures at the tangent space $T_{p}$ for each spacetime point $p$. The first one, determined by $\left\{k^{\mu}\right\}$ (normal vector), is called the ``normal cone'', while the other, spanned by $\left\{q^{\mu}\right\}$ (tangent vector) is called the ``characteristic cone''. Since the objects $\hat g^{\mu\nu}$ and $\hat g_{\mu\nu}$ vary from point to point, the two cones may acquire different geometrical/topological properties. In this sense, the propagation properties of the nonlinear theory described by the action (\ref{action}) is determined by two types of cone fields.


\section{Algebraic structure of $\phi^{\mu\nu}$}
\label{class}
The effective metric given in Eqn.(\ref{in1})
depends of the object $\phi_{\mu\nu}$. We shall see next that the properties of the mixed object $\phi^{\mu}_{\ ~\nu}$ not only determine the type of the quasi-linear partial differential equation (\ref{pde}), but also the basic properties of the effective geometry $\hat g^{\mu\nu}$. it is easily seen that Eqn. (\ref{pde}) can be rewritten as 
\begin{equation}\label{equ}
\hat g^{\mu\nu}\phi_{,\mu,\nu}+...=0
\end{equation}
where the dots stand for terms involving only first order derivatives of and algebraic terms in $\phi$. Thus, the principal part of the differential equation is entirely determined by the contravariant components of the effetive metric. The eigendirections of $\phi^{\mu}_{\ ~\nu}$ are given by the equation
\begin{equation}
\Phi\vec\xi=\lambda\vec\xi,
\end{equation}
where the components of the matrix $\Phi$ are defined as $\Phi\doteq \phi^{\mu}_{\phantom a\nu}$ and $\vec\xi$ is an eigenvector associated to the eigenvalue $\lambda$. The characteristic equation, given by
$ {\rm det}(\Phi-\lambda\textbf{1})=0,
$
can be evaluated using the expression ${\rm Tr}\;\Phi^{n}=W^{n}$ (where $n>0$ is a natural number) 
and the Cayley-Hamilton theorem. The result is 
\begin{equation}
\lambda^{3}(\lambda-W)=0,
\end{equation}
showing that there are two different possible eigenvalues $\lambda_{1}=W$ and $\lambda_{2}=0$ with multiplicities one and three, respectively. 

In the following we shall use $\left\{X^{\mu}\right\}$ and $\left\{Y^{\mu}\right\}$ as the sets of eigenvectors associated with the eigenvalues $\lambda_{1}$ and $\lambda_{2}$ respectively, thus yelding
\begin{equation}
(\phi_{,\nu}X^{\nu})\phi^{,\mu}=WX^{\mu},\label{eigen1}
\end{equation}
\begin{equation}
(\phi_{,\nu}Y^{\nu})\phi^{,\mu}=0 \label{eigen2}.
\end{equation}
From Eqn.(\ref{eigen1}) it is seen that, for $W\neq0$, there exists a single direction associated to the eigenvalue $\lambda_{1}$. This direction is determined by the field gradient $\phi^{,\mu}$. We will consider next the three possible cases according to the nature of the gradient $\phi_{,\mu}$, \textit{i.e.}  timelike, lightlike and spacelike. We shall be specially interested in the null eigenvectors of $\Phi$ since,
as it will be shown below, they are associated with special types of propagation. 

\begin{itemize}
\item{When $W>0$, Eqn.(\ref{eigen1}) guarantees that a given $X^{\mu}$ is proportional to the gradient $\phi^{,\mu}$, implying that it is timelike \textit{i.e.} $X^{\mu}X_{\mu}>0$. From Eqn.(\ref{eigen2}) we see that $Y^{\mu}$ and $X^{\mu}$ aree orthogonal. Thus, any vector contained in the three-space $\mathcal H$ orthogonal to the gradient is an eigenvector $Y^{\mu}$ with eigenvalue $\lambda_{2}$. This leads to a degenerate structure where an infinite set of vectors corresponds to the same $\lambda$. By choosing a linearly independent triad of orthonormal vectors $Y^{\mu}_{(i)}$ ($i=1,2,3$) in the space $\mathcal H$ together with the first eigenvector $X^{\mu}$ ($X^{\mu}X_{\mu}=1$) as a basis, it is possible to diagonalize the matrix $\Phi$. Note that, in this special case, the algebraic structure of $\Phi$ is such that it does not admit any null eigenvector.}\label{1}

\item{In the case of a lightlike gradient ($W=0$), $\lambda_{1}$ and $\lambda_{2}$ are both zero. A straightforward calculation shows that the sets $\left\{X^{\mu}\right\}$ and $\left\{Y^{\mu}\right\}$ are both determined by all possible vectors orthogonal to $\phi^{,\mu}$. This unique set constitutes a three dimensional vector space. It does not admit any timelike element and it has only one null principal direction, which is proportional to the gradient $\phi^{,\mu}$ itself. It is important to emphasize that, although a symmetric linear operator can always be diagonalizad in a space with positive definite metric, this is not the case in a space with Lorentzian signature. In fact, because a linearly independent set of eigenvectors of $\Phi$ does not exist in the case under scrutiny, it is not possible to diagonalize the matrix $\Phi$ \footnote{Note that a similar situation occurs in the study of the energy-momentum tensor of null electromagnetic fields (see, for instance \cite{Synge,Landau})}.\label{2} 

\item{When $W<0$, a simple inspection of Eqn.(\ref{eigen1}) shows that $X^{\mu}X_{\mu}<0$. In turn, Eqn. (\ref{eigen2}) implies that the set $\left\{Y^{\mu}\right\}$ is determined by vectors orthogonal to a given $X^{\mu}$. Neverthless, the resultant three-space $\mathcal H$ has Lorentzian signature (because it is orthogonal to a spacelike vector). Hence, it contains timelike directions. By choosing a linearly independent triad of orthonormal vectors $Y^{\mu}_{(i)}$ ($i=0,1,2$) in $\mathcal H$ such that $g_{\mu\nu}Y^{\mu}_{(0)}Y^{\nu}_{(0)}=1$ and $g_{\mu\nu}Y^{\mu}_{(j)}Y^{\nu}_{(j)}=-1$ ($j=1,2$) together with $X^{\mu}$ ($X^{\mu}X_{\mu}=1$) as a basis, it is possible to bring $\Phi$ to a diagonal form. 
}}\label{3}\\
\end{itemize}
Notice that although this last case seems to be similar to the case in which $W>0$, the Lorentzian signature of $\mathcal H$ leads to the existence of null eigenvectors. In fact, there exists an infinite number of null eigeinvectors of $\Phi$ that belong to the two-dimensional subspace of $\mathcal H$ defined by the equation $Y^{\mu}Y_{\mu}=0$. We shall see that these vectors will play an important role in the propagation properties in a background field with a spacelike character.

\section{The cone field and the hyperbolic regime}
\label{cone}
We now restrict to the hyperbolic case, in which the three-dimensional surface given by Eqn.$(\ref{cones})$ coincides with the characteristics of Eqn.(\ref{eom}). The different types of $\phi^{\mu}_{\ ~\nu}$ determine whether the new characteristics intersect the usual light cones of Minkowski geometry. These intersections are of interest since along them the propagation of the excitations is unaffected by the background field. Let us study them in detail, begining with a simple lemma related to the structure of the characteristic cone at a given point and its relation to the background light cone.
\\
\textit{Lemma:} At a given point P of the background spacetime $\mathcal M$, a characteristic vector $q^{\mu}$ intersects the light cone of the background only if it is a null eigenvector of the tensor $\phi^{\mu}_{\phantom a\nu}$.\\
\textit{Proof:} To intersect the light cone, the characteristic vector $q^{\mu}$ must satisfy at P the relations
\begin{equation}
g_{\mu\nu}q^{\mu}q^{\nu}=0, \quad\quad 2\call_{WW}(\phi_{,\mu}q^{\mu})^2=0.
\end{equation}
Now, the second relation holds only if $\call_{WW}=0$ or $\phi_{,\mu}q^{\mu}=0$. The first case is trivial since if $\call_{WW}=0$, the new characteristic coincides the light cones of the linear theory. We thus impose that $\call_{WW}\neq0$ and $\phi_{,\mu}q^{\mu}=0$, \textit{i.e} $q^{\mu}$ is a null eigenvector of $\Phi$ and belongs both to the background light cone and the characteristic cone.
\\\\
We investigate next what types of characteristics are allowed for each type of field. The definitions $\nabla^{\mu}\phi\equiv(\dot\phi,-\vec\nabla\phi)$ and $q^{\mu}\equiv(q^{0},\vec q)$ will be used below.

\subsection{Time-like field}
This is by far the most simple and familiar situation since the symmetries of the observed universe according to the standard cosmological model impose that $W>0$. In an orthogonal frame such that $\nabla^{\mu}\phi=\sqrt{W}\delta^{\mu}_{0}$ we obtain that the new ``cones" are described by 
\begin{equation}
c_{s}^{-2}\left[(q^{1})^2+(q^{2})^{2}+(q^{3})^{2})\right]=(q^{0})^{2},
\label{cone1}
\end{equation}
where the ''velocity of sound'' $c_s$ is given by
$$
c_{s}^{-2}\equiv\frac{\vec{q}.\vec{q}}{(q^{0})^2}=\frac{\call_{W}}{\call_{W}+2\call_{WW}W}
$$
Eqn.(\ref{cone1}) shows that the two-dimensional surfaces obtained by setting $q^{0}=$const. are centered spheres, which may be interpreted as the wave fronts for the scalar field as seen by an observer at rest with the background field $\phi_{0}$. We note also that the excitations travel with the same velocity in all directions with respect to this frame. Neverthless, because in this case $\phi^{\mu}_{\phantom a\nu}$ does not admit null eigenvectors, according to the lemma above, there are no intersections between the new characteristic cone and the background light cones (this is also seen directly from Eqn.(\ref{cone1})). Finally, note that there will be faster than light propagation if $\call_{WW}<0$.

\subsection{Null field}

Although we have not been able to find in the literature a discussion of null fields
in this context, they are important mainly because of two reasons: 1) they describe transition  regions of the background solutions that admit both timelike and spacelike gradients $\nabla^{\mu}\phi$; 2) they represent exact plane wave solutions of the field equations when the Lagrangian is not an explicit function of $\phi$.\\ 
With respect to an orthogonal basis, where $\nabla^{\mu}\phi=A(1,1,0,0)$, we obtain after a straightforward calculation the covariant effective metric
\begin{equation}
\label{matrix_g_mu_nu_nf}
\hat g_{\mu\nu}=\frac{1}{\mathcal{L}_{W}^{2}}\left(
\begin{array}{cccc}
\mathcal{L}_W-2\mathcal{L}_{WW}A^{2}&2\mathcal{L}_{W}&0&0\\
2\mathcal{L}_{W}&-\mathcal{L}_W-2\mathcal{L}_{WW}A^{2}&0&0\\
0&0&-\mathcal{L}_{W}&0\\
0&0&0&-\mathcal{L}_{W}
\end{array}
\right)
\end{equation}
Using Eqn.(\ref{in1}), we obtain that the characteristic cone satisfies
\begin{equation}
\left(\call_{W}+2\call_{WW}A^{2}\right)(q^{1})^2+\call_{W}\left[(q^{2})^{2}+(q^{3})^{2}\right]-4\call_{WW}A^{2}q^{0}q^{1}=(\call_{W}-2\call_{WW}A^{2})(q^{0})^{2}.
\end{equation}
It follows that only vectors of the type $q^{\mu}\propto\nabla^{\mu}\phi$ satisfiy this equation and at the same time are null with respect to the background metric. This means that the characteristics of the nonlinear scalar field intersect the light cone of the background spacetime along one single spacetime direction given by the gradient $\nabla^{\mu}\phi$.    

\subsection{Space-like field}

By choosing an orthonormal basis with the third axis pointing along $\vec\nabla\phi$ it is possible to write the equation of the characteristic surface at a point $P$ as
\begin{equation}
\left[(q^{1})^2+(q^{2})^{2}\right]+c_s^2(q^{3})^{2}=(q^{0})^{2}
\end{equation}
with $w<0$. Hence, there exists different velocities of propagation associated to each axis. We obtain
\begin{equation}
c_{s,1}^2=1,\quad c_{s,2}^2=1, \quad c_{s,3}^2=c_s^2.
\label{ort}
\end{equation}
Thus, along directions orthogonal to the spacelike gradient $\nabla^{\mu}\phi$ the scalar excitations propagate with the velocity of light. If $c_s^2>1$ there exist at least one direction where tachionic trajectories are allowed. In general, if $\vec q$ denotes an arbitrary direction in the three-space the ``velocity of sound"
is
\begin{equation}
v_{s}^{2}(\theta)=\left\{1+(c_s^2-1)\;{\rm cos}^{2}\theta\right\},
\end{equation}
where $\theta$ is the angle between $\vec q$ and $\vec\nabla\phi$.

\subsection{Examples}

We shall illustrate next the classification given above with two examples coming from a generalized  Dirac-Born-Infeld (DBI) scalar model, with action given by
\beq
S = \int \left[ 1-(1+U(\phi))\:\sqrt{1-W}\;\right] d^4x,
\label{adop}
\eeq
where a mass scale has been set equal to one, 
and the function $U(\phi)$ acts as a potential in the low-energy limit \cite{doppel}.

\subsubsection{Static background ($W<0$)}

As shown in \cite{doppel}, given a canonical 
scalar field theory with a positive semidefinite potential $V(\phi)$ which enables domain wall solutions,
there exists a choice for $U(\phi)$ in the theory defined by Eqn.(\ref{adop}), given by 
\beq
U(\phi) = -1+\sqrt{1+2V(\phi)},
\label{pot}
\eeq 
which yields domain walls of the same field profile and energy
density. Then, by choosing $V(\phi) = \frac\lambda 4 (\phi^2-v^2)^2$,
the theory described by the action given in Eqn.(\ref{adop}) with $U$ given by Eqn.(\ref{pot}) 
has a domain wall solution of the form
\beq
\phi(z) = v\tanh \left(\sqrt{\frac\lambda 2}\;v (z-z_0)\right).
\label{wall}
\eeq
The gradient of this background solution is space-like, and the associated effective metric can be calculated from
Eqn.(\ref{effmet}). The relevant components are 
\beq
\hat g^{tt} =\frac{1+U(\phi)}{2}\left(1-W\right)^{-1/2},
\eeq
\beq
\hat g^{zz} = -\frac{1+U(\phi)}{2}\left(1-W\right)^{-3/2},
\eeq
with $U(\phi) = -1+\sqrt{1+ \frac\lambda 2 (\phi^2-v^2)^2}$, and 
$W = -\frac{\lambda v^4}{2}\cosh^{-4}\left(v\sqrt{\frac\lambda 2}t\right)$.
The lightcone structure associated to this effective metric is shown in 
Fig.(\ref{static}).
\begin{figure}[H]
       \centering  
       \includegraphics[scale=.5]{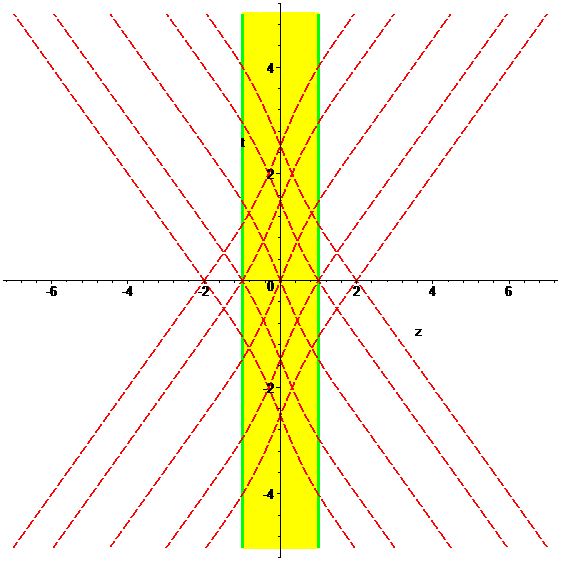}
       \caption{The lightcones of the high-energy perturbations for the DBI scalar model are given by the intersections of the geodesics of the effective metric. The domain wall - in yellow - was plotted by assuming that the field has a linear dependence with $z$ in the region of spacetime where $\phi$ is appreciably different fromits limiting values. The values $v=\sqrt 2$ and $\lambda = 1$ were used in the plot.}
       \label{static}
\end{figure}
It is seen that the lightcones concide with those of the Minkowskian geometry far from the wall, 
and get thinner inside it, signaling that high-energy perturbations propagate sub-luminally there.
We also see that the propagation in the directions $x$ and $y$ is not affected by the wall, in accordance with Eqn.(\ref{ort}).  
The time spent by a given perturbation inside the wall depends on the parameters of the potential, 
but the wall cannot trap the high-energy perturbations (as also shown in \cite{doppel}).


\subsubsection{Dynamical background ($W>0$)}

If the background field $\phi_0$ is such that $\phi_0 = \phi_0 (t)$, the equation of motion (\ref{eom}) 
can be written as
\beq
\left[\ddot \phi (\call_W + 2\call_{WW}\dot\phi^2)-\half \call_\phi\right]_0=0.
\label{tdeom}
\eeq
Time-dependent solutions can only be valid approximately, as in the case of 
the zero mode of a vibrating string
with fixed ends, if we restric to the region near the center of the string. 
By imposing that the solution be oscillatory, 
\beq
\phi_0 = \phi_c \sin(\omega t),
\label{osc}
\eeq
(with $\phi_c$ a constant) in Eqn.(\ref{tdeom}) and using the lagrangian given in Eqn.(\ref{adop}),
we get that this solution is possible for the potential
\beq
V (\phi ) = -1+V_0\exp\left\{-\frac{\omega^2}{4} (\phi_c^2-\phi^2)\left[2+\omega^2(\phi_c^2-\phi^2)\right]\right\},
\label{potv}
\eeq
where $V_0$ is a constant.
This potential is regular for every value of $\phi$, 
displays two symmetric peaks separated by a minimum, and goes to zero
for large absolute values of the field. 

For the oscillatory background solution given by Eqn.(\ref{osc}), the effective metric is 
\beq
\hat g_{00} = \left.\frac{1+V}{2}\frac{1+2W}{(1-W)^{3/2}}\right|_0,\;\;\;\;\;\;\;\;\;\;
\hat g_{11} = \hat g_{22} = \hat g_{33} =-\left.\frac{1+V}{2\sqrt{1-W}}\right|_0.
\label{emosc}
\eeq
Using the potential in Eqn.(\ref{potv}), the following equation is obtained for the rays:
\beq
\frac{dx}{dt} = \pm \sqrt{\frac{1+2\omega^2\phi_0^2\cos^2(\omega t)}{1-\omega^2\phi_0^2\cos^2(\omega t)}},
\eeq
Notice that there is a nonzero conformal factor in the metric given in Eqn.(\ref{emosc}), which cancels out in this equation. The following plot shows some of the rays.
\begin{figure}[H]
       \centering  
       \includegraphics[scale=.5]{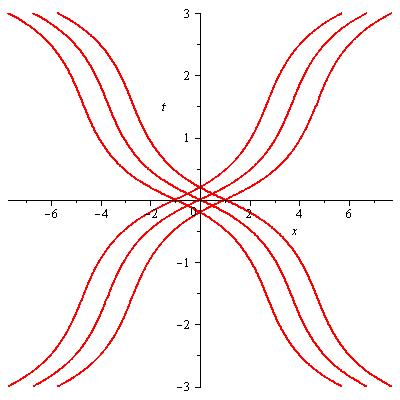}
       \caption{Lightcones of the high-energy perturbations for the DBI theory in the case of time-dependent solutions. The values $\omega = 1$ and $\phi_c = 1/\sqrt 2$ were used in the plot.}
       \label{dynamic}
\end{figure}
We see that along with the modification of the lightcones, the oscillatory background
focuses and defocuses the rays in a periodic manner and, in accordance with Eqn.(\ref{cone1}), 
the lightcones oscillate bewteen those of Minkowski spacetime and
the ``external'' light cones, which set the maximal velocity for super-luminal propagation at each spacetime point.
\section{Effective signatures and background field types} 
\label{effe}
We have mentioned before that the type of the quasi-linear equation (\ref{equ}) depends on the signature of the matrix $\hat g^{\mu\nu}$, and the inverse of this object is responsible for the causal properties of the theory given by the action (\ref{action}). This is because the characteristic cone is entirely determined by the set of vectors $\left\{q^{\mu}\right\}$ that satisfy the relation given in Eqn.(\ref{disp2}). It will be shown in this section 
that by lifting the requirement of hyperbolicity of the field equations
\footnote{The condition $\call_{W}>0$ being still valid.}, 
nonlinearities allow different signatures for the effective metric. We shall also show that the admissible effective signatures depend strongly on the background field type \footnote{It is important to note that in some cases the dynamics of a field theory can be uniquely defined even in the absence of hyperbolicity. See for instance the series of papers by Wald and Ishibashi where the evolution of a scalar field obeying the Klein-Gordon equation is defined uniquely in non-globally hyperbolic spacetimes \cite{ishi}.}.

Let us begin by the investigation of some properties of the object $\hat g^{\mu\nu}$ by means of its relationship with the tensor $\Phi$ defined above. It is convenient to introduce the mixed object $\hat\textbf{G}$
\begin{equation}
\hat\textbf{G}\equiv\hat{g}^{\mu\alpha}g_{\alpha\nu}=\mathcal{L}_{W}\textbf{1}+2\mathcal{L}_{WW}\Phi.
\end{equation} 
At a given point of the manifold $\mathcal{M}$, $\hat\textbf{G}$ can be thought as a linear map of the tangent space $T_{p}$ onto itself. It is immediate to see that if $\vec\xi$ is an eigenvector of $\Phi$ with eigenvalue $\lambda$ it will also be an eigenvector of $\hat\textbf{G}$ with eigenvalue $\gamma=\mathcal{L}_{W}+2\mathcal{L}_{WW}\lambda$. The two possibilities are, thus
\begin{equation}\label{eigen}
\gamma_1=\mathcal{L}_{W}+2\mathcal{L}_{WW}W,\quad\quad\quad\quad\gamma_2=\mathcal{L}_{W}.
\end{equation}
As discussed in the previous sections, there are three possible types of algebraic structures for the object $\phi^{\mu}_{\phantom a\nu}$. We shall analyze next 
the structure of $\hat\textbf{G}$ for each of these. 

\subsection{Time-like field ($W>0$)}

This is the simplest situation from the algebraic point of view. In terms of the orthonormal basis $X^{\mu}$ and $Y^{\mu}_{(i)}$ defined in case 1 of Section \ref{1}, $\hat\textbf{G}$ can be written as: 
\begin{equation}
\hat\textbf{G}={\rm diag}(\gamma_{1},\gamma_{2},\gamma_{2},\gamma_{2}),
\end{equation}
with the eigenvalues $\gamma_1$ and $\gamma_2$ given by Eqn.(\ref{eigen}).
 
By choosing at a given point of $\mathcal M$ a local cartesian system, \textit{i.e} $g^{\mu\nu}(P)=\eta^{\mu\nu}$, the contravariant effective metric $\hat g^{\mu\nu}$ can be written as  
\begin{equation}\label{dis}
\hat g^{\mu\nu}={\rm diag}(\gamma_{1},-\gamma_{2},-\gamma_{2},-\gamma_{2})
\end{equation}
Thus, depending on the Lagrangian density $\call(\phi,w)$ under investigation, the following possibilities for the signature are allowed in the case $W>0$:
\begin{displaymath}
\begin{array}{lllll}
\hline\hline \gamma_{1}&\gamma_{2}&{\rm PDE\; type}& {\rm signature} & {\rm n_{eff}}\\ 
\hline + & > &{\rm Hyperbolic}&+---&4\\ 
\hline 0& >&{\rm Parabolic}& \;---&3\\
\hline - & >&{\rm Elliptic}&----&4\\
\hline
\end{array}
\end{displaymath}
Notice that from the diagonalized form of the metric 
given in Eqn.(\ref{dis}) it follows that are three possible types of PDE regimes. Since we are considering only stable theories, these types are determined by the sign of $\gamma_{1}$. We see also that the differential equation for the field is hyperbolic if $\gamma_{1}$ is positive and elliptic if $\gamma_{1}$ is negative. The first situation leads to an effective spacetime with a Lorentzian signature while the second leads to an Euclidean effective spacetime, endowed with not three but four spacelike directions.

If $\gamma_{1}$ vanishes, the equation becomes of parabolic type. It seems that in this special regime the effective dimension ${\rm n_{eff}}$ is reduced and  the emergent spacetime is three-dimensional. This suggests that the parabolic regime may be related to prohibited regions for the excitation propagation.   

\subsection{Null field}

We start with a Lorentz frame at a given point of Minkowski spacetime and with a field gradient such that $\phi^{,\mu}=(1,1,0,0)$. With respect to this frame, $\hat\textbf{G}$ has the form
\begin{equation}
\label{matrix_g_mu_nu_0}
\hat\textbf{G}=\left(
\begin{array}{cccc}
\mathcal{L}_W+2\mathcal{L}_{WW}&-2\mathcal{L}_{WW}&0&0\\
2\mathcal{L}_{WW}&\mathcal{L}_W-2\mathcal{L}_{WW}&0&0\\
0&0&\mathcal{L}_W&0\\
0&0&0&\mathcal{L}_W
\end{array}
\right)
\end{equation}
The non-diagonal terms are present because the matrix $\Phi$ does not admit a complete set of eigenvectors. Thus, we cannot obtain a diagonal representation of the effective metric $\hat g^{\mu\nu}$ simply raising the operator $\hat\textbf{G}$ with the background metric $\eta^{\mu\nu}$. Neverthless, we can perform a coordinate transformation such that the resultant effective metric is of the form 
\begin{equation}
\hat g^{\mu\nu}={\rm diag}(\gamma_{+},\gamma_{-},-\gamma_{2},-\gamma_{2})
\end{equation}
with $\gamma_{\pm}\equiv 2\call_{WW}\pm\sqrt{4\call_{WW}^{2}+\call_{W}^2}$ and $\gamma_{2}=\call_{W}$. Now, because $\gamma_{+}>0$ and $\gamma_{-}<0$, the signature depends only on the sign of $\gamma_{2}$. Because we are assuming that $\gamma_{2}>0$ we now we have only one possibility, summarized as follows:

\begin{displaymath}
\begin{array}{lllll}
\hline\hline \gamma_{2}& {\rm PDE\; type}& {\rm signature} & {\rm n_{eff}}\\ 
\hline > &{\rm Hyperbolic} &+---&4\\ 
\hline
\end{array}
\end{displaymath}
Thus, if the background field is such that its gradient is lightlike everywhere, both the partial differential equation and the effective spacetime signature are globally hyperbolic. This analogue spacetimes are particularly important if one is interested to study the interaction of nonlinear scalar waves. 
\subsection{Space-like field}

This field type imply an interesting spectra of possibilities for the effective metric. Because the eigenvector $X^{\mu}$ is spacelike, there exist an orthonormal coordinate system at any point $P$ such that $\hat\textbf{G}$ has the diagonal form  
\begin{equation}
\hat\textbf{G}={\rm diag}(\gamma_{2},\gamma_{2},\gamma_{2},\gamma_{1})
\end{equation}
with $\gamma_{i}$, $(i=1,2)$ given by Eqn.(\ref{eigen}). Multiplication of $\hat\textbf{G}$ with $\eta^{\mu\nu}$ leads to the following components for the contravariant effective metric
$\hat g^{\mu\nu}={\rm diag}(\gamma_{2},-\gamma_{2},-\gamma_{2},-\gamma_{1}).
$ and to the following combinations:
\begin{displaymath}
\begin{array}{lllll}
\hline\hline \gamma_{1} & \gamma_{2}& {\rm PDE\;type} & {\rm signature} & {\rm n_{eff}}\\ 
\hline > & > &{\rm Hyperbolic}&+---&4\\ 
\hline =& >&{\rm Parabolic}& +--&3\\
\hline < & >&{\rm UltraHyp.}&++--&4\\
\hline
\end{array}
\end{displaymath}
In this case there are also three possibilities, governed by the sign of the eigenvalue $\gamma_{1}$. If $\gamma_{1}$ is positive, again we have a hyperbolic regime for the PDE while if it vanishes we obtain a parabolic regime. From this point of view the situation is quite similar to that one studied in the case of a time-like field. Nevertheless when the eigenvalue is negative we do not obtain an elliptic equation anymore. In contrast, we obtain an ultrahyperbolic equation governing the field. Thus, the effective spacetime is such that its signature is of the form $(++--)$, indicating that we should have an additional emergent timelike dimension. Although from the mathematical point of view, the issue of satisfactory initial or bounday conditions for ultrahyerbolic equations is not understood, physics with two time dimensions has been studied in several articles \cite{2t}.

To close this section we would like to remark that the condition for the hiperbolicity of the PDE is the same for all three types of background fields. In fact, if $\gamma_{1}>0$ it is automatically guaranteed that the theory is governed by hyperbolic equations independently of the background, a condition that was already pointed out in \cite{aha}. 


\section{Discussion}

Nonlinear scalar field theories have been repeatedly used to model physical phenomena,
particularly in Cosmology and Astrophysics. We have discussed in detail the features of the propagation of the high-energy perturbations in these theories, which is governed by the 
effective metric. We have shown that the effective metric can be classified in three classes, 
according to whether the gradient of the background scalar field is spacelike, null, or timelike. 
Examples of some of these types were examined, and the results were shown to be in agreeement with the classification. \\
In the second part of the paper, we have discussed the possibility of signature changes
of the effective metric, due to the change in character of the differential equation
obeyed by the background field. Depending on the different possibilities
for the gradient of the backgroundd field, it was shown that the partial differential equation for the background field can be hyperbolic, parabolic, elliptic, or ultra-hyperbolic. Some interestig cases arise, such as an effective spacetime with one dimension less than the background, or with 
two time-like dimensions, and will be studied in a future publication.

\section*{Acknowledgements}

SEPB would like to acknowledge support from CNPQ, and ICRANet-Pescara. Both authors acknowledge 
support from FAPERJ.

\end{document}